\documentclass[fleqn,usenatbib]{mnras}

\usepackage{newtxtext,newtxmath}

\usepackage[T1]{fontenc}

\DeclareRobustCommand{\VAN}[3]{#2}
\let\VANthebibliography\thebibliography
\def\thebibliography{\DeclareRobustCommand{\VAN}[3]{##3}\VANthebibliography}

\usepackage{graphicx}	
\usepackage{amsmath}	

\usepackage{xcolor} 
\usepackage{siunitx}    
\DeclareSIUnit\gauss{G}
\DeclareSIUnit\msun{M_\odot}
\DeclareSIUnit\year{yr}

\newcommand{\V}{V1460~Her}

\title[Optical detection of the rapid spin of V1460 Her]{Optical detection of the rapidly spinning white dwarf in V1460 Her}

\author[Pelisoli et al.]{
Ingrid Pelisoli,$^{1}$\thanks{E-mail: ingrid.pelisoli@warwick.ac.uk}
T.~R.\ Marsh,$^{1}$
R.~P.\ Ashley,$^{2}$
Pasi Hakala,$^{3}$
A.\ Aungwerojwit,$^{4}$
K.\ Burdge,$^{5}$
\newauthor
E. Breedt,$^{6}$
A.~J. Brown,$^{7}$
K. Chanthorn,$^{8}$
V.~S. Dhillon,$^{7,9}$
M.~J. Dyer,$^{7}$
M.~J. Green,$^{10}$
P. Kerry,$^{7}$
\newauthor
S.~P. Littlefair,$^{7}$
S.~G. Parsons,$^{7}$
D.~I. Sahman,$^{7}$
J.~F. Wild,$^{7}$
S. Yotthanathong$^{8}$
\\
$^{1}$Department of Physics, University of Warwick, Gibbet Hill Road, Coventry, CV4 7AL, UK\\
$^{2}$Isaac Newton Group of Telescopes, Apartado de Correos 321, Santa Cruz de La Palma, E-38700, Spain\\
$^{3}$Finnish Centre for Astronomy with ESO (FINCA), Quantum, University of Turku, FI-20014, Finland\\
$^{4}$Department of Physics, Faculty of Science, Naresuan University, Phitsanulok, 65000, Thailand\\
$^{5}$Division of Physics, Mathematics and Astronomy, California Institute of Technology, Pasadena, CA 91125, USA\\
$^{6}$Institute of Astronomy, University of Cambridge, Madingley Road, Cambridge CB3 0HA, UK\\
$^{7}$Department of Physics and Astronomy, Hicks Building, The University of Sheffield, Sheffield, S3 7RH, UK\\
$^{8}$National Astronomical Research Institute of Thailand, 260 Moo 4, T. Donkaew, A. Maerim, Chiangmai, 50180, Thailand\\
$^{9}$Instituto de Astrof\'{i}sica de Canarias, E-38205 La Laguna, Tenerife, Spain\\
$^{10}$Department of Astrophysics, School of Physics and Astronomy, Tel Aviv University, Tel Aviv 6997801, Israel
}

\date{Accepted XXX. Received YYY; in original form ZZZ}

\pubyear{2021}

\begin{document}

\label{firstpage}
\pagerange{\pageref{firstpage}--\pageref{lastpage}}
\maketitle

\begin{abstract}
Accreting magnetic white dwarfs offer an opportunity to understand the interplay between spin-up and spin-down torques in binary systems. Monitoring of the white dwarf spin may reveal whether the white dwarf spin is currently in a state of near-equilibrium, or of uni-directional evolution towards longer or shorter periods, reflecting the recent history of the system and providing constraints for evolutionary models. This makes the monitoring of the spin history of magnetic white dwarfs of high interest. In this paper we report the results of a campaign of follow-up optical photometry to detect and track the \SI{39}{\sec} white dwarf spin pulses recently discovered in \textit{Hubble Space Telescope} data of the cataclysmic variable V1460~Her. We find the spin pulsations to be present in $g$-band photometry at a typical amplitude of \SI{0.4}{\percent}. Under favourable observing conditions, the spin signal is detectable using 2-meter class telescopes. We measured pulse-arrival times for all our observations, which allowed us to derive a precise ephemeris for the white dwarf spin. We have also derived an orbital modulation correction that can be applied to the measurements. With our limited baseline of just over four years, we detect no evidence yet for spin-up or spin-down of the white dwarf, obtaining a lower limit of $|P/\dot{P}| > 4\times10^{7}$~years, which is already 4 to 8 times longer than the timescales measured in two other cataclysmic variable systems containing rapidly rotating white dwarfs, AE~Aqr and
AR~Sco.

\end{abstract}




\section{Introduction}

The spin rates of accreting white dwarfs in cataclysmic variable binary stars are determined by competing mechanisms which add or extract angular momentum. On the one hand,
accreting material of high specific angular momentum drives the white dwarf to spin faster. The concentrated internal structures of white dwarfs, together with their shrinkage in response to increased mass, means that
only $\sim$ \SI{0.1}{\msun} of added matter is needed in principle to bring a white dwarf close to its few second maximal breakup spin rate \citep{LivioPringle1998}.
However, there are no systems known to be very close to this limit, presumably because white dwarfs can also lose angular momentum in three ways. First, white dwarfs accreting hydrogen-rich material undergo thermonuclear runaways once \SI{1e-6}{\msun} to \SI{1e-4}{\msun} of matter has accumulated, causing their envelopes to expand \citep{Paczynski1978, Wolf2013}. Core-envelope coupling during these phases can be expected to slow the white dwarfs' spin rates \citep{LivioPringle1998}. Second, magnetic white dwarfs couple through their fields to the accretion disc or to the binary companion, causing their spin rates to slow. In the limit of very strong fields their spins lock to the binary orbit, as seen in the polar class of accreting white dwarfs, also known as AM~Her stars \citep{Joss1979,Campbell1983}. A third possible mechanism is through tides,
which are likely to be important for ultra-compact orbits and during double white dwarf mergers \citep{Fuller2012}.

Magnetic white dwarfs are of particular interest because they reveal their spin periods through photometric variations induced by spots, which result from accretion rate asymmetries caused by their fields. This allows the measurement of extremely precise spin periods to reveal the competing effects of accretion versus magnetic drag on a year by year basis \citep{Patterson1984}. This plays out in the intermediate polar (IP) class of system. IPs have spin periods of order minutes to tens of minutes \citep{Patterson1994}. These are rates much slower than breakup, but faster than their binary periods, reflecting a quasi-equilibrium between spin-up and magnetic drag that is continuously adjusting to a fluctuating accretion rate \citep{Patterson1984, Kennedy2016, Littlefield2020}

Not all systems exist in  quasi-equilibrium. One of the first recognised cataclysmic variables, AE~Aquarii, shows remarkable flaring behaviour that is thought to be driven by magnetically-propelled material exiting the system \citep{Wynn1997}. AE~Aqr was unique in this behaviour amongst the thousands of known systems until the recent discovery of LAMOST-J024048.51+195226.9 \citep{Thorstensen2020, Garnavich2021}. The white dwarf in AE~Aqr has a spin period of \SI{33}{\sec} \citep{Patterson1979} and is spinning down on a short \SI{e7}{\year} timescale \citep{deJager1994} and it is thought that little or no accretion takes place. Instead the system is now in a spin-powered state which powers not just the flares, but also a broad synchrotron spectrum extending to radio frequencies \citep{Bookbinder1987}. AE~Aqr is not in equilibrium, but is instead being observed in what appears to be a relatively brief evolutionary phase compared to the several billion year lifetimes of cataclysmic variables. A similar, possibly more advanced, state has been reached by the system AR~Sco, which, like AE~Aqr, is also a strong synchrotron source \citep{Marsh2016}. The white dwarf in AR~Sco, whose spin period is \SI{118}{\sec}, is spinning down on a timescale of \SI{5e6}{\year} \citep{Stiller2018, Gaibor2020}, and in this case the system appears to be entirely detached \citep{Marsh2016,GarnavichLittlefield2021}, with no accretion and no evidence for propeller-induced flaring activity \citep{Littlefield2017}. 

It is not yet clear how systems like AE~Aqr and AR~Sco achieved their current states. Although their magnetic field strengths have not been directly measured, the absence of accretion discs and the rates of spin down have led to estimates ranging from \SI{50}{\mega\gauss} to \SI{200}{\mega\gauss} \citep{Ikhsanov1998, Katz2017}. White dwarfs with fields this high are hard to spin up to the short periods seen in AE~Aqr and AR~Sco, because very high accretion rates are needed to compress the magnetosphere down to a radius at which the Keplerian orbital period in the disc is this short. While there are signs of a high rate of accretion in AE~Aqr's recent past \citep{Schenker2002}, there is no similar evidence in the case of AR~Sco. Perhaps the field estimates are simply far too large \citep{Lyutikov2020}, but that would still leave an open question of why these two systems in particular have lost their discs.

A radically different hypothesis for AR~Sco has been recently put forward by \citet{Schreiber2021}. They proposed that the white dwarf in AR~Sco only became magnetic as a result of a crystallisation- and rotation-driven dynamo, similar to the mechanism thought to be at work in planets and low-mass stars. In their model, which was derived from an explanation for white dwarf magnetism initially put forward by \citet{Isern2017}, the carbon-oxygen white dwarf in a post common envelope binary with a main sequence star is originally non-magnetic. As the system evolves towards shorter orbital periods, the white dwarf cools and the companion evolves towards a Roche-lobe filling star. When that happens, the binary becomes a cataclysmic variable, and accretion spins up the white dwarf, whose core might be crystallising. If that is the case, conditions for a dynamo -- strong density stratification and convection -- are met, generating a magnetic field. If the field is strong enough, the disc may be disrupted and connection with the secondary star field will provide a synchronising torque on the white dwarf spin. Again, if the field is strong enough, the rapid transfer of spin angular momentum into the orbit may cause the binary to detach and mass transfer to cease. AR~Sco might be an example of such a system. This provides a natural explanation for the high incidence of magnetism amongst cataclysmic variable stars \citep[36\%,][]{Pala2020}, together with a near total absence of magnetism amongst their progenitor systems \citep{Liebert2005, Liebert2015, Parsons2021}. Importantly, it sidesteps the problem of spinning up a highly magnetic white dwarf.

\citet{Schreiber2021}'s hypothesis makes accreting white dwarfs of short spin period of very high interest, as they could be systems whose fields have only recently emerged, and they might now be in a state of rapid spin down. Whether this is the case, or whether they are in fact in a state of quasi-equilibrium, possible for relatively weak fields at short spin periods, can only be established through observation.

One such short period white dwarf has been recently-discovered in V1460~Her \citep{Ashley2020}. V1460~Her is an eclipsing cataclysmic variable with a \SI{4.99}{\hour} orbital period and an over-luminous K5-type donor star. It is the third fastest spinning white dwarf known amongst the cataclysmic variables, with a spin period of \SI{38.875\pm 0.005}{\sec} \citep{Ashley2020}. Amongst cataclysmic variable stars, only the white dwarfs in AE~Aqr (\SI{33}{\sec}) and CTCV~J2056-3014 \citep[\SI{29.6}{\sec},][]{Oliveira2020} spin faster, although the X-ray binary HD~49798 contains a $P_\mathrm{spin} = \SI{13}{\sec}$ compact component that might be a white dwarf \citep{Israel1997,Mereghetti2011}. An evolved companion is another unusual characteristic among CVs that is shared by \V\ and AE~Aqr. Only $5\pm3$\% of CVs are found to have evolved companions \citep{Pala2020}, which is at odds with the theoretical prediction of 30\% \citep{Schenker2002}.

Since discovering the rapid pulsations betraying the presence of the white dwarf spin in V1460~Her's HST UV data \citep{Ashley2020}, we have obtained multiple epochs of high-speed optical photometry in an effort to detect and monitor the spin. In this paper we present the results of this campaign aimed at establishing the spin history of the magnetic white dwarf V1460~Her since its recent discovery.

\section{Observations}

Our follow-up observations of \V\ were carried out at the \mbox{2.0-m} Liverpool Telescope (LT), the 2.56-m Nordic Optical Telescope (NOT), the 4.2-m William Herschel Telescope (WHT), the Palomar 200-inch telescope, and the 2.4-m Thai National Telescope (TNT). The details of our observing runs are shown in Table~\ref{tab:obs_log}. 

At the LT, we used the optical wide-field camera IO:O with a 2x2 binning to reduce the readout time. Observations were taken with the Sloan $g$ filter. Given a rather slow readout time of \SI{18}{\sec}, we chose exposure times of 10--\SI{12}{\sec} to avoid too much loss of signal, even though the resulting cadence of 28.6--\SI{30.6}{\sec} was sub-Nyquist for the spin signal in \V.

The WHT runs used the QHY CMOS detector at prime  focus with a windowed configuration aimed at minimising readout time. For the two initial runs in February 2021, the Astronomik B filter, which has a similar throughput to Sloan $g$, was used. For the following two runs, in May 2021, the Sloan $g$ filter was used. The exposure time was set to \SI{5}{\sec}, which resulted on a cadence of \SI{7.8}{\sec} accounting for the readout time of \SI{2.8}{\sec}.

At the TNT, the high-speed camera ULTRASPEC \citep{Dhillon2014} was employed, with a $g$ filter and exposure time varying between 4.8 and 9.6~sec depending on the observing conditions, with only 15~msec lost between exposures.

NOT observations were executed with ALFOSC, a multimode imager/spectrograph. We used a subwindow binned by a factor of two yielding a 2 sec dead time. A $g$ filter was used, except for observations starting on 2020-05-30, which used a $U$ filter. The exposure time for both g and $U$ band data was 3 sec.

The Caltech HIgh-speed Multi-color camERA \citep[CHIMERA,][]{chimera} provided data from the 200-inch Hale telescope at Palomar observatory. The $g$ filter was used in the blue channel. In the red, the K-type star companion dominates the light, diluting the signal from the white dwarf. Therefore, only the blue channel data are used here. The CCD was operated using the 0.1 MHz conventional amplifier with 1 second exposures with 2x2 binning, and used frame transfer to effectively eliminate time lost due to reading out between exposures.

All our observations were bias subtracted and flat-field corrected. Observation times were corrected to Barycentric Julian Date (BJD) in the Barycentric Dynamical Time (TDB) reference system. For each image, differential aperture photometry was carried out. We have used a variable aperture size, set to scale with the seeing measured from a point-spread function (PSF) fit, to accommodate variations in the atmospheric conditions.
We chose as our main comparison {\it Gaia} EDR3~1382561212513336064, a nearby star with similar colour to \V\ and, in particular, showing low astrometric excess noise and Renormalised Unit Weight Error (RUWE) in {\it Gaia}, which are proxies for variability \citep{Belokurov2020}. For one of our TNT observations,
using this comparison star yielded poor results. In this case, we have used instead a brighter comparison star, TYC~3068-855-1. Using this bright comparison is not possible for most of our observations, as it is often saturated. For the NOT runs, SDSS~J162123.10+441241.6 was used as the comparison star.

\begin{table*}
	\centering
	\caption{Journal of observations}
	\label{tab:obs_log}
	\begin{tabular}{cccccc}
		\hline
		Run number & Telescope & Start date (TDB) & Duration (min) & Cadence (s) & Filter\\
		\hline
		1 & TNT & 2020-03-11 21:15:54.281 & 99.6 & 4.9 & $g$\\
        2 & TNT & 2020-03-13 21:06:52.664 & 53.5 & 4.8 & $g$ \\
        3 & TNT & 2020-03-14 21:07:02.065 & 122.0 & 4.3 & $g$ \\
        4 & TNT & 2020-03-15 20:59:57.109 & 76.7 & 7.3 & $g$ \\
        5 & TNT & 2020-03-23 22:30:16.836 & 19.8 & 7.9 & $g$ \\
        6 & NOT & 2020-05-02 02:58:03.805 & 63.7 & 5.0 & $g$ \\
        7 & NOT & 2020-05-30 01:33:11.873 & 76.8 & 6.5 & $U$ \\
        8 & NOT & 2020-07-30 23:33:21.091 & 63.4 & 5.0 & $g$ \\
        9 & Palomar & 2020-07-31 04:29:04.918 & 33.6 & 1.0 & $g$ \\
        10 & LT & 2021-01-29 05:36:56.940 & 49.5 & 30.6 & $g$ \\
        11 & LT & 2021-02-14 05:24:35.987 & 49.5 & 30.6 & $g$ \\
        12 & WHT & 2021-02-14 04:51:50.774 & 103.9 & 7.8 & $B$ \\
        13 & WHT & 2021-02-17 03:13:28.577 & 153.2 & 7.8 & $B$ \\
        14 & TNT & 2021-02-28 21:53:35.618 & 72.1 & 4.8 & $g$ \\
        15 & TNT & 2021-03-01 21:57:58.907 & 71.8 & 5.0 & $g$ \\
        16 & TNT & 2021-03-02 21:46:42.073 & 80.2 & 4.8 & $g$ \\
        17 & LT & 2021-03-03 03:02:07.962 & 49.5 & 30.6 & $g$ \\
        18 & TNT & 2021-03-09 21:27:09.409 & 94.1 & 5.4 & $g$ \\
        19 & LT & 2021-03-18 04:06:46.081 & 49.1 & 28.6 & $g$ \\
        20 & TNT & 2021-03-20 20:40:35.052 & 132.6 & 6.0 & $g$ \\
        21 & TNT & 2021-03-20 20:40:35.052 & 132.6 & 6.0 & $g$ \\
        22 & TNT & 2021-03-31 21:31:34.211 & 77.8 & 9.6 & $g$ \\
        23 & TNT & 2021-04-01 21:22:14.809 & 81.8 & 9.6 & $g$ \\
        24 & LT & 2021-04-02 05:09:16.121 & 49.3 & 28.6 & $g$ \\
        25 & TNT & 2021-04-09 18:32:39.911 & 235.6 & 8.9 & $g$ \\
        26 & TNT & 2021-04-10 20:31:04.867 & 116.0 & 8.9 & $g$ \\
        27 & TNT & 2021-04-18 20:35:01.170 & 120.4 & 4.8 & $g$ \\
        28 & TNT & 2021-04-19 18:52:00.371 & 58.2 & 4.8 & $g$\\
        29 & NOT & 2021-05-01 02:25:01.479 & 62.5 & 6.0 & $g$ \\
        30 & TNT & 2021-05-02 20:18:49.762 & 121.3 & 4.8 & $g$ \\
        31 & TNT & 2021-05-03 19:15:18.515 & 178.2 & 4.8 & $g$ \\
        32 & TNT & 2021-05-04 20:18:16.380 & 21.0 & 9.6 & $g$\\
        33 & TNT & 2021-05-05 20:05:27.439 & 112.9 & 9.6 & $g$\\
        34 & WHT & 2021-05-19 22:05:18.250 & 440.0 & 7.8 & $g$\\
        35 & WHT & 2021-05-21 02:32:53.930 & 169.1 & 7.8 & $g$\\
		\hline
	\end{tabular}
\end{table*}

\section{Results}

\subsection{Detection of the pulsations}
\label{sec:detection} 

\begin{figure}
	\includegraphics[width=\columnwidth]{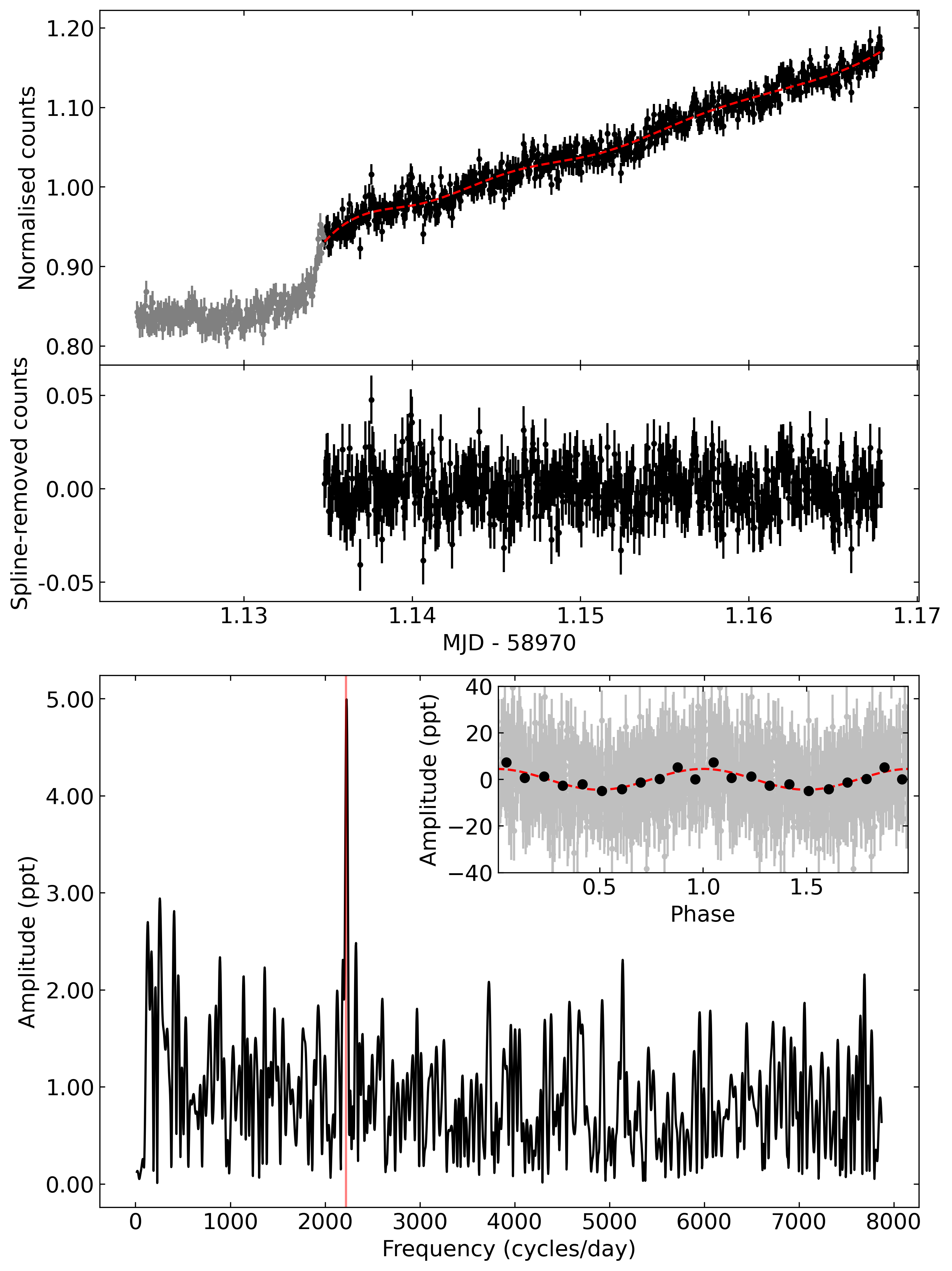}
    \caption{Top panel shows the original light curve, taken at the NOT on 2020-05-02. Data started being taken during an eclipse, shown in grey. Only the data shown in black was used for fitting a spline, shown as a red dashed line. The spline-subtracted data is shown in the middle panel. The amplitude spectrum of these data is shown in the bottom panel, with the spin period derived by \citet{Ashley2020} shown as a vertical red line. Amplitudes are shown in parts per thousand (ppt). The inset in the bottom panel shows the spline-subtracted data (in grey) folded on the spin period, with the cosine fit shown as a red dashed line. The black points show an average every 50 points to aid visualisation.}
    \label{fig:fit_ft}
\end{figure}

\begin{figure*}
    \centering
	\includegraphics[width=\textwidth]{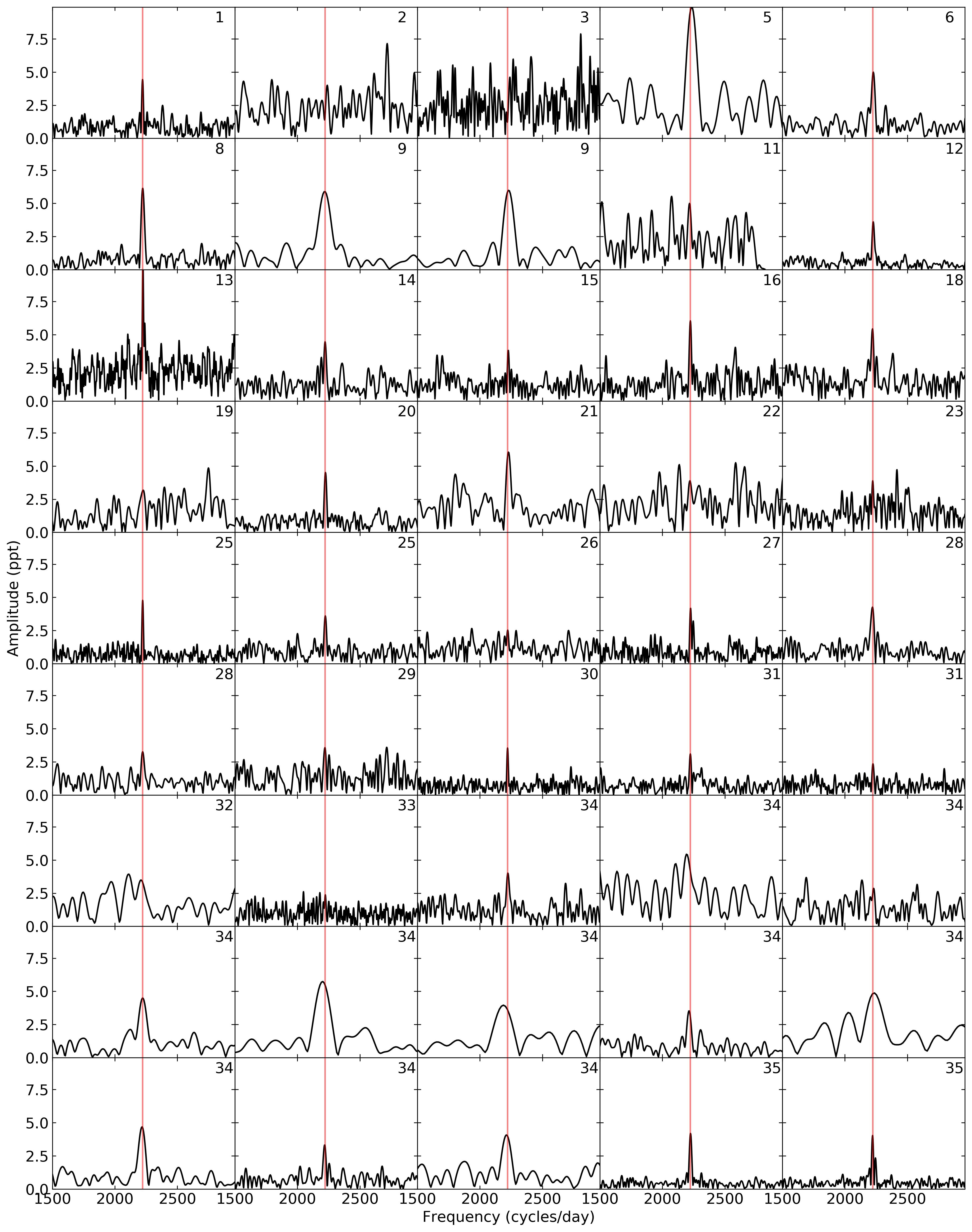}
    \caption{Amplitude spectra corresponding to the derived pulse times shown in Table~\ref{tab:t0_cyc}. The run number is shown at the top right corner, and the spin period obtained by \citet{Ashley2020} is marked by a vertical red line. As described in the text, long runs were split into smaller time intervals, therefore more than one panel is shown for a few runs, each panel corresponding to one of the time intervals.}
    \label{fig:all_ft}
\end{figure*}

We computed the amplitude spectrum for each separate run, following subtraction of a spline to remove orbital-related variations. Data taken while the white dwarf was eclipsed were masked. An example of this procedure is shown in Fig.~\ref{fig:fit_ft}, and the amplitude spectra of all runs is shown in Fig.~\ref{fig:all_ft}. The spin pulsation signal was always detected in our WHT runs, and is clear in our CHIMERA run as well. It is also detected in the three $g$-band NOT observations, though with high uncertainty for one of them (due to bright Moon conditions), but not in the $U$-band run. For the TNT runs, the signal is detected in all but three runs, which were affected by detector pickup noise and/or poor observing conditions (clouds and/or bright Moon). In the case of the Liverpool Telescope observations, detection is only marginally possible owing to the slow detector readout time along with the consequent sub-Nyquist sampling and loss of signal. We detected the signal with an uncertainty smaller than 30\% in only one of our five LT runs.

\subsection{Measurement of pulse times}
\label{sec:measureT0}

We measured a single time and uncertainty for each discrete run representing the time of peak flux $T_0$ by fitting a cosine with a period fixed to \citet{Ashley2020}'s measurement (see Fig.~\ref{fig:fit_ft} inset), but with an arbitrary zeropoint in time. The fits were carried out to the spline-subtracted data described in Section~\ref{sec:detection}. The zeropoint times from such measurements are uncertain by arbitrary offsets of integer multiples of the period. However, any error in the period used for the fit will be amplified if offsets are chosen that move the selected time far from the data upon which it is based. Therefore we selected values close to the mid-time of each dataset in every case. Runs of a length significant compared to the \SI{4.99}{\hour} orbital period were split into sub-sections in order to be sensitive to possible orbital modulation, which is expected if the periodic signal contains orbital side-bands as is often the case in IPs \citep[e.g.][]{Kennedy2016, Patterson2020}. Table~\ref{tab:t0_cyc} shows the measured times, along with cycle number counts whose derivation we detail in the next section. Uncertainties are given by the standard deviation after a thousand bootstrapping iterations, in which data points were sampled allowing for repeated values and then refitted.

\begin{table}
	\centering
	\caption{Measured pulse times, with respective uncertainties, and the derived cycle count. As noted in the text, some runs were split into smaller time intervals and thus contributed to more than one $T_0$ measurement. Run number corresponds to that shown in Table~\ref{tab:obs_log}, with the exception of the previously existing HST run from \citet{Ashley2020} (labelled as 0).}
	\label{tab:t0_cyc}
    \begin{tabular}{cccc}
        \hline
        Run number & $T_0$ (BJD TDB) & $\sigma_{T0}$ & Cycle \\
        \hline
        0 &	57814.9463288 & 0.0000070 & -2383862 \\
        0 &	57815.0043813 & 0.0000059 & -2383733 \\
        0 &	57815.0835622 & 0.0000052 & -2383558 \\
        1 &	58919.9207031 & 0.0000128 & 70872 \\
        2 &	58921.8984927 & 0.0002144 & 75265 \\
        3 &	58922.9163339 & 0.0002217 & 77526 \\
        4 &	58923.8926391 & 0.0005235 & 79695 \\
        5 &	58931.9447283 & 0.0000133 & 97583 \\
        6 &	58971.1515796 & 0.0000113 & 184682 \\
        7 &	58999.0953122 & 0.0004916 & 246760 \\
        8 &	59061.0036053 & 0.0000085 & 384292 \\
        9 &	59061.1925888 & 0.0000093 & 384711 \\
        9 &	59061.2038316 & 0.0000071 & 384736 \\
        10 & 59243.2513696 & 0.0005262 & 789161 \\
        11 & 59259.2425308 & 0.0000252 & 824685 \\
        12 & 59259.2465427 & 0.0000125 & 824694 \\
        13 & 59262.1911550 & 0.0000088 & 831236 \\
        14 & 59273.9344339 & 0.0000152 & 857324 \\
        15 & 59274.9407973 & 0.0000382 & 859560 \\
        16 & 59275.9355367 & 0.0000145 & 861770 \\
        17 & 59276.1437909 & 0.0005388 & 862232 \\
        18 & 59282.9359955 & 0.0000166 & 877321 \\
        19 & 59291.1902899 & 0.0002677 & 895658 \\
        20 & 59293.8860530 & 0.0000110 & 901647 \\
        21 & 59293.9418275 & 0.0000150 & 901771 \\
        22 & 59304.9117311 & 0.0000745 & 926141 \\
        23 & 59305.9190805 & 0.0000258 & 928379 \\
        24 & 59306.2321825 & 0.0012266 & 929074 \\
        25 & 59313.8161005 & 0.0000109 & 945922 \\
        25 & 59313.9096758 & 0.0000160 & 946130 \\
        26 & 59314.8774209 & 0.0000443 & 948280 \\
        27 & 59322.8937543 & 0.0000143 & 966089 \\
        28 & 59323.8065797 & 0.0000135 & 968117 \\
        28 & 59323.9149702 & 0.0000162 & 968357 \\
        29 & 59335.1223806 & 0.0001813 & 993255 \\
        30 & 59336.8887346 & 0.0000121 & 997179 \\
        31 & 59337.8330519 & 0.0000144 & 999277 \\
        31 & 59337.8947119 & 0.0000199 & 999414 \\
        32 & 59338.8534525 & 0.0000594 & 1001544 \\
        33 & 59339.8801360 & 0.0000240 & 1003824 \\
        34 & 59353.9416886 & 0.0000122 & 1035063 \\
        34 & 59353.9691594 & 0.0000121 & 1035124 \\
        34 & 59354.0204276 & 0.0000122 & 1035238 \\
        34 & 59354.0487706 & 0.0000117 & 1035300 \\
        34 & 59354.0613222 & 0.0000122 & 1035328 \\
        34 & 59354.0689996 & 0.0000111 & 1035345 \\
        34 & 59354.0847275 & 0.0000119 & 1035380 \\
        34 & 59354.1009802 & 0.0000117 & 1035416 \\
        34 & 59354.1135250 & 0.0000120 & 1035444 \\
        34 & 59354.1441434 & 0.0000120 & 1035512 \\
        34 & 59354.2134852 & 0.0000123 & 1035666 \\
        35 & 59355.1316792 & 0.0000091 & 1037706 \\
        35 & 59355.2036616 & 0.0000108 & 1037866 \\
        \hline
    \end{tabular}
\end{table}

\subsection{Fixing the cycle counts}

\begin{figure}
    \centering
	\includegraphics[width=\columnwidth]{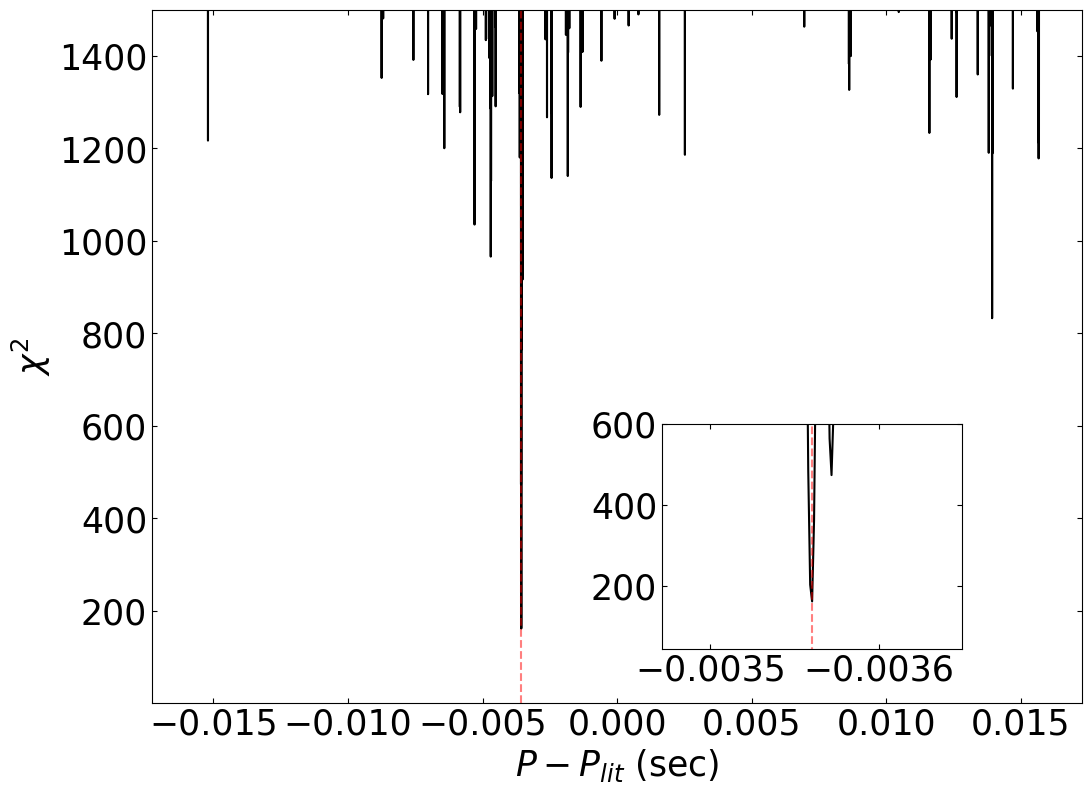}
    \caption{Minimum value of $\chi^2$ of a linear ephemeris fit over a range of periods spanning five times the uncertainty reported in \citet{Ashley2020}. The period range is shown in terms of the difference to the period in \citet{Ashley2020}, $P_{lit}$. Our best period is indicated by the vertical dashed line. The inset shows a zoom around the best period, showing that there are no competing period aliases with similar $\chi^2$ values.}
    \label{fig:Pchi}
\end{figure}

We calculated integer pulse cycle numbers for all data over a finely-spaced grid covering two dimensions representing adopted spin period and spin phase offset, the two unknowns of the problem. We searched over a set of periods centred on the previously determined period of \SI{38.875\pm0.005}{\sec} from \citet{Ashley2020}, extending over an interval of $\pm 5\sigma$ on either side, where $\sigma$ is \citet{Ashley2020}'s uncertainty estimate. The phase offset grid extended from 0 to 1, i.e. over an entire cycle. Any given period/phase offset pair within the grid leads to a set of integer cycle counts which can then be used to fit a linear ephemeris to the data from which a $\chi^2$ value results. For any fixed period, it is important to optimise over the phase offset since it is not known in advance and it can result in cycle numbers flipping by plus or minus a cycle, hence our use of a 2D search grid. Fig.~\ref{fig:Pchi} 
shows the resulting $\chi^2$ as a function of period, after selection of the minimum $\chi^2$ over all phase offsets for each period.

This is the stage at which one can tell if the observations are sufficient to fix the cycle counts uniquely. If they are not, there will be multiple aliases with similar minimum $\chi^2$ values. Any with comparable values of $\chi^2$ need to be regarded as potential candidate periods, with more data being required to distinguish between them. We find that the second deepest minimum has a $\chi^2$ more than 300 larger than our favoured period (see Fig.~\ref{fig:Pchi}), and still $>100$ after scaling uncertainties to give reduced $\chi^2 = 1$. This is comfortably above the threshold difference of 10 found to yield reliable periods by \citet{MRueda2003} in their study of sdB orbital periods\footnote{The reasoning is that the probability of a period is dominated by the term exp($-\chi^2/2$), therefore a difference of more than 10 implies that the second-best alias is at least exp$(5) \simeq 150$ times less probable than the best.}. This absence of any competing aliases shows that there is only one viable period and that the cycle counts are uniquely determined. The location of the best period within less than $2\sigma$ of \citet{Ashley2020}'s favoured value is further support of the proposed solution.

In order to define the epoch of cycle number 0, we performed a linear ephemeris fit for different $T_0$ values, separated by an integer number of cycles and spanning the time interval of our observations. We monitored the resulting covariance between $T_0$ and period in order to find the location of the minimum. We elected the value that minimised this covariance as $T_0$. This results in the following ephemeris for the time of maximum brightness:
\begin{equation}
    \textrm{BMJD(TDB)} = 58888.0183797(17) + 0.0004498987920(12) E,
    \label{eq:eph1}
\end{equation}
where E is the cycle number. The time scale is TDB, corrected to the barycentre of the solar system, expressed as a Modified Julian Day number (MJD = JD-2400000.5). Uncertainties were determined from a thousand Monte-Carlo runs of the linear ephemeris fit, resampling the derived times of maxima each time according to their uncertainties.

\subsection{Modulation with orbital period}

\begin{figure}
	\includegraphics[width=\columnwidth]{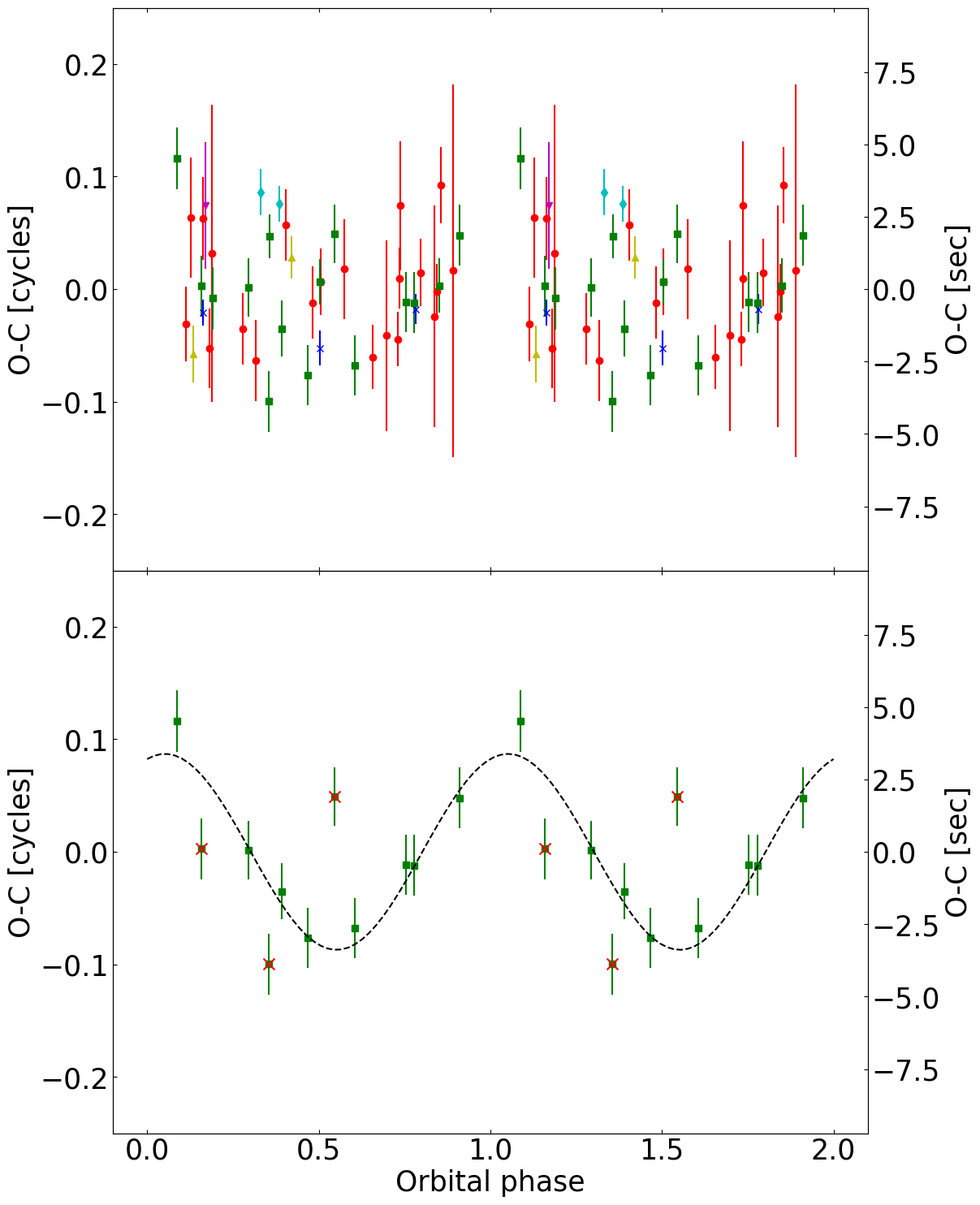}
    \caption{Top panel: Pulse timing delay as a function of the orbital phase. \citet{Ashley2020}'s HST data are shown as blue crosses, TNT data as red circles, WHT data as green squares, LT data as magenta up-side-down triangles, NOT data as yellow triangles, and CHIMERA data as cyan diamonds. Points with uncertainty larger than 30\% of the spin period are not shown. A hint of orbital modulation can be noticed, with pulses presenting a slight delay around phases zero and 1.0, and a shorter arrival time around phase 0.5. Bottom panel: Same as top panel, but showing only the continuous 7.3~hour WHT run. Removing the effect of instrument sensitivity and minimising weather effects makes the orbital modulation clear. The dashed line shows a least-squares fit to model this modulation and calculate a correction. Points marked by crosses were not used in the fit.}
    \label{fig:orbit}
\end{figure}

In order to identify whether the spin signal presents any orbital modulation, we have calculated the orbital phase for each of our measurements using the orbital ephemeris derived by \citet{Ashley2020}. We have subtracted the calculated (C) cycle count from the observed (O) value for each epoch to inspect the O-C diagram for any hints of dependence on the orbital phase. The top panel of Fig.~\ref{fig:orbit} shows the result for all our observing runs. The highly variable size of the errorbars, which are strongly dependent on the instrument and observing conditions, make it hard to achieve any conclusion regarding orbital modulation.

To minimise the effect of observing conditions and remove the effect of different instruments, we have repeated the same exercise for a single WHT run with a duration of 7.3~hours. Given the high quality of the WHT data, the spin period can often be detected within a short time interval. To optimise the number of detections for this run, our procedure was to start with a small chunk of data spanning five minutes, and compute the value of $T_0$ as described in Section~\ref{sec:measureT0}. If the obtained value had an uncertainty smaller than 3\% of the spin period, it was accepted; otherwise the time interval was increased by two minutes. This was repeated iteratively until the desired uncertainty was achieved, or the time interval reached an hour. Once a value was accepted, the same procedure was repeated for the next chunk of data, starting with the first value after the previous time interval, i.e. we allowed for no overlap between the data ranges used for each detection. The result is shown in the bottom panel of Fig.~\ref{fig:orbit}. With this more homogeneous dataset, the orbital modulation becomes clear.

We can therefore estimate a correction to derived $T_0$ values depending on the orbital phase in order to remove any effects of modulation. This modulation is not solely the result of the white dwarf orbital motion, which can explain at most $\sim 1.8$~sec out of the observed range of $\sim 7$~sec, but is likely largely caused by the effect of beat signals slightly shifting the period peak. As shown in Fig.~\ref{fig:orbit}, we calculate the modulation correction by fitting a sinusoidal curve to the WHT data. We obtained:
\begin{equation}
    T_0{}_{\textrm{corr}} = 3.40(43)~\textrm{sec} \times \sin[2 \pi \varphi_{\textrm{orb}} + 1.25(6)]
    \label{eq:T0cor}
\end{equation}
where $\varphi_{\textrm{orb}}$ is the orbital phase of each measurement, and the phase constant shown is in radians. Uncertainties were obtained by bootstrapping. We note that this correction is not applied for the times shown in Table~\ref{tab:t0_cyc}. Applying this correction, we have re-derived our ephemeris, obtaining:
\begin{equation}
    \textrm{BMJD(TDB)} = 58888.0183511(17) +  0.0004498988053(12) E.
    \label{eq:eph2}
\end{equation}

\begin{figure}
	\includegraphics[width=\columnwidth]{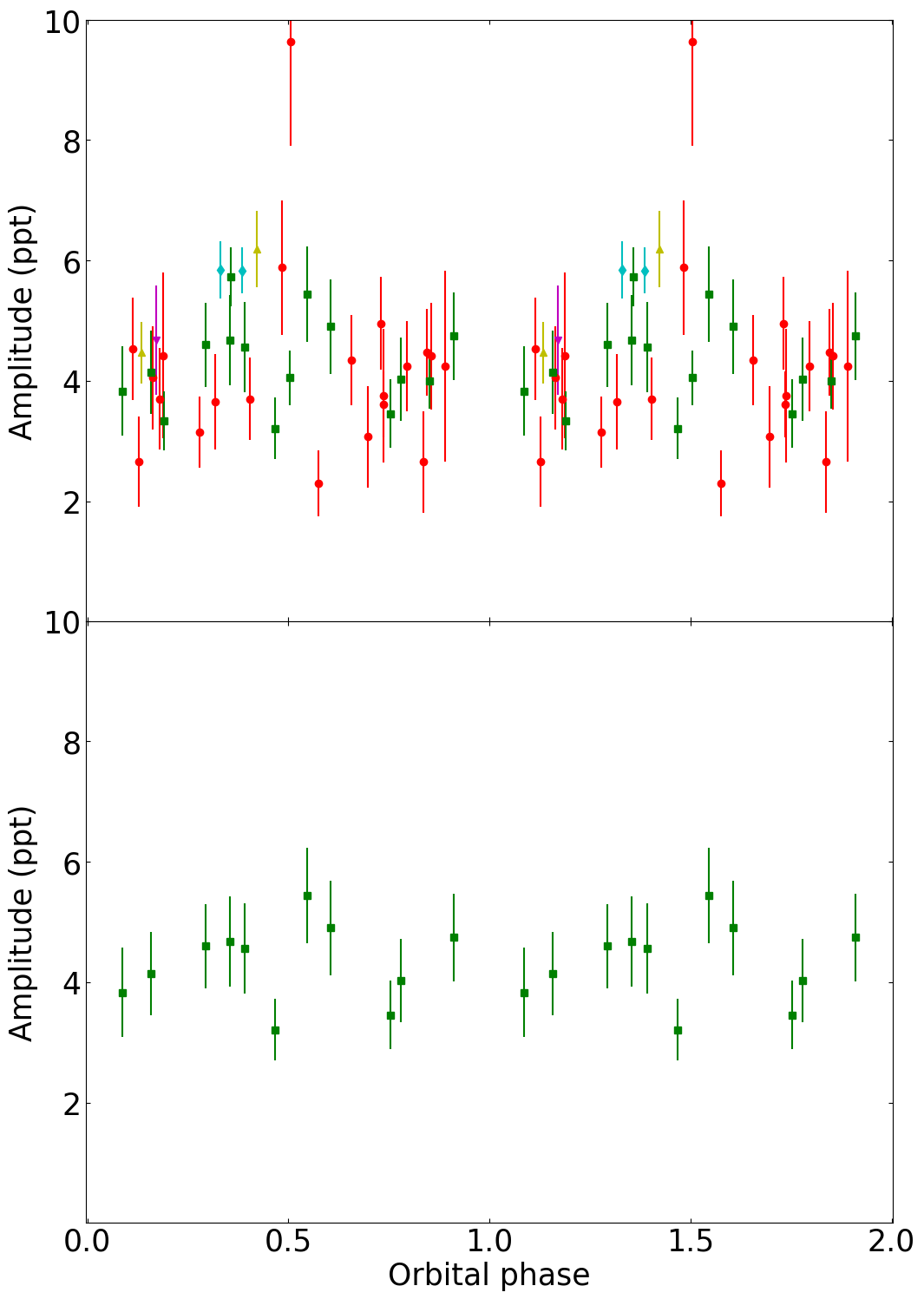}
    \caption{Top panel: Amplitude of the spin pulsation as a function of the orbital phase. Symbols are the sames as in Fig.~\ref{fig:orbit}. There is no evidence of amplitude variability, and the scatter is consistent with Gaussian noise. Bottom panel: Same as top panel, for only the continuous 7.3~hour WHT run.}
    \label{fig:orbit_amp}
\end{figure}

Similarly, we can also investigate any orbital modulation of the spin amplitude. Fig.~\ref{fig:orbit_amp} shows the estimated pulse amplitude as a function of orbital phase. Unlike the pulse times, the amplitudes show no dependence on the orbital phase, and the observed scatter is consistent with a Gaussian distribution according to a normality test.

\subsection{The long-term spin behaviour of the white dwarf}

With a baseline of only just over four years, and with essentially just three observing "seasons" in place, it is early days to constrain the spin evolution of the white dwarf in \V. However, already we have some sensitivity to the short evolutionary timescales seen in the systems AE~Aqr and AR~Sco discussed in the introduction.

\begin{figure}
	\includegraphics[width=\columnwidth]{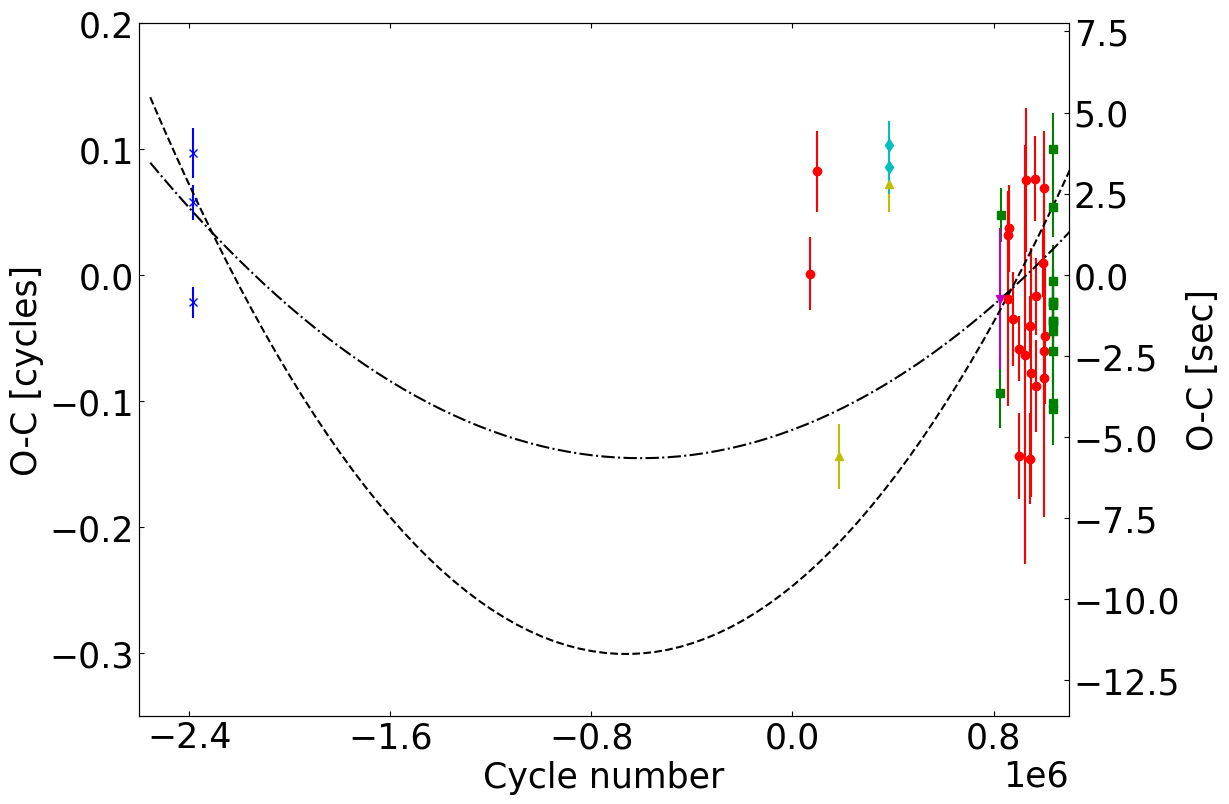}
    \caption{Pulse timing delay as a function of time. Symbols are the same as for Fig.~\ref{fig:orbit}. The dashed line shows a best fit to the data assuming the spin-down timescale of AR~Sco ($P/\dot{P} = 5\times 10^6$~yr), and the dot-dashed line shows the same for AE~Aqr ($P/\dot{P} = 1\times 10^7$~yr). Although sparsely sampled in the first three years of coverage, the measurements around cycle number 0 suggest that V1460~Her does not have spin variations of comparable magnitude.}
    \label{fig:ominusc}
\end{figure}

After applying the correction given by Eq.~\ref{eq:T0cor} to our derived pulse times, we have inspected the O-C diagram as a function of cycle number as shown in Fig.~\ref{fig:ominusc}. We show for comparison the expected O-C behaviour for a spin-down of $P/\dot{P} = 5 \times 10^6$~yr \citep[as estimated for AR~Sco,][]{Stiller2018}, and  for $10^7$~yr \citep[that of AE~Aqr,][]{deJager1994}. From our current dataset, there does not seem to be any indication of spin variability for \V, at least in timescales comparable to the these two similar systems. The O-C measurements are instead scattered around zero (see Fig.~\ref{fig:ominusc}).

In order to place a lower limit on the evolutionary timescale of the spin period, we performed a quadratic fit to the ephemeris, corrected for the orbital modulation, using the Markov-Chain Monte Carlo method, implemented with {\tt emcee} \citep{emcee}. As expected, the quadratic term is poorly constrained and consistent with zero. Adopting the 1\% percentile as the lower limit, we obtain $|P/\dot{P}| > 4\times10^{7}$~yr. With the detection of the optical pulsations and an ephemeris secure over a timescale of years, it should be possible to improve upon this limit rapidly in the future since any quadratic signal grows with the square of the baseline. However, this timescale, which is longer than is seen in either AE~Aqr or AR~Sco, suggests that \V\ may be in a state of quasi-equilibrium as opposed to a short-lived state of spin down.


\section{Summary \& Conclusions}

We have detected the spin pulsations of \V\ in the optical for the first time. This is particularly important for long-term monitoring of the spin, as optical detection implies that observations can be carried out with ground-based telescopes. Our data shows that the spin can be detected even for 2-meter class telescopes like the TNT, provided that the observing conditions are favourable.

We have executed 35 observing runs over more than one year. Combining our observations with the HST data from \citet{Ashley2020}, we obtained a total of 44 detections of the spin with timing uncertainty better than 30\%, given that some runs allowed for more than one measurement. This allowed for a much more precise estimate of the ephemeris (Eq.~\ref{eq:eph1}) compared to the values of \citet{Ashley2020}.

We have also identified an orbital modulation of the pulse arrival times. Modelling this effect, we have derived a correction (Eq.~\ref{eq:T0cor}, which we applied to our derived times to present improved ephemeris (Eq.~\ref{eq:eph2}), and to investigate the long-term behaviour of the spin frequency. Although we believe we see signs of modulation of the pulse phase with orbital period, we note that the amplitude of the correction is less than 0.1 spin cycles, so our conclusions are not qualitatively affected even if we apply no correction.

With the current baseline of just over four years, albeit sparsely sampled at the start, 
we find no evidence for any change in the spin period of \V, and place a lower
limit on the timescale of spin period changes of $|P/\dot{P}| > 4\times10^{7}$~years.
This suggests that \V\ could currently be in a state of quasi-equilibrium caused by a balance between accretion spin-up, and spin-down triggered by magnetic torque. Measuring the spin variability in accreting fast-spinning white dwarfs can ultimately enable us to probe the origin of magnetic fields in cataclysmic variables \citep{Schreiber2021}, providing constraints on the evolution of interacting compact binaries. Therefore, following the main result of this work that the spin pulsations can be detected in the optical, we encourage continuous monitoring of this system in order to confirm the quasi-equilibrium behaviour, or possibly measure the spin change with a longer baseline than presented here.

\section*{Acknowledgements}

We thank the referee John Thorstensen for reviewing our manuscript. IP and TRM acknowledge support from the UK's Science and Technology Facilities Council (STFC), grant ST/T000406/1 and from the Leverhulme Trust. AA acknowledges support from Thailand Science Research and Innovation (TSRI) grant FRB640025 contract no. R2564B006. VSD is supported by the STFC grant ST/V000853/1.  SGP acknowledges the support of a STFC Ernest Rutherford Fellowship.

This work has made use of data obtained at the Thai National Observatory on Doi Inthanon, operated by NARIT. The data presented here were partially obtained with ALFOSC, which is provided by the Instituto de Astrofisica de Andalucia (IAA) under a joint agreement with the University of Copenhagen and NOT.

\section*{Data Availability}

All data analysed in this work can be made available upon reasonable request to the authors.
 



\bibliographystyle{mnras}
\bibliography{v1460her} 








\bsp	
\label{lastpage}
\end{document}